\documentclass[aps,prd,twocolumn,amsmath,amssymb,superscriptaddress,floatfix,nofootinbib]{revtex4-1}
\usepackage{color}

\usepackage{graphicx,psfrag} 
\usepackage{hyperref}
\usepackage{verbatim}
\newcommand{\newc}{\newcommand}
\newc{\gsim}{\lower.7ex\hbox{$\;\stackrel{\textstyle>}{\sim}\;$}}
\newc{\lsim}{\lower.7ex\hbox{$\;\stackrel{\textstyle<}{\sim}\;$}}
\newc{\gev}{\,{\rm GeV}}
\newc{\mev}{\,{\rm MeV}}
\newc{\ev}{\,{\rm eV}}
\newc{\kev}{\,{\rm keV}}
\newc{\tev}{\,{\rm TeV}}

\def\tr{\mathop{\rm tr}}

\newc{\mz}{M_Z}
\newc{\mpl}{M_*}
\newc{\mw}{m_{\rm weak}}
\newc{\nr}[1]{N^c_R{}_{#1}}

\newcommand{\second}{{\, {\rm s}}}

\newcommand{\cm}{{\, {\rm cm}}}

\newcommand{\eV}{{\, {\rm eV}}}

\newcommand{\GeV}{{\, {\rm GeV}}}

\newcommand{\sun}{\odot}

\newc{\qcd}{\,{\rm QCD}}

\usepackage{amsmath}
\def\beq{\begin{equation}}
\def\eeq{\end{equation}}
\def\bea{\begin{eqnarray}}
\def\eea{\end{eqnarray}}
\def\bitem{\begin{itemize}}
\def\eitem{\end{itemize}}
\newc{\ie}{{\it i.e.}}          \newc{\etal}{{\it et al.}}
\newc{\eg}{{\it e.g.}}          \newc{\etc}{{\it etc.}}
\newc{\cf}{{\it c.f.}}
\def\bar#1{\overline{#1}}

\def\inv{^{\raise.15ex\hbox{${\scriptscriptstyle -}$}\kern-.05em 1}}
\def\lbar{{\lower.35ex\hbox{$\mathchar'26$}\mkern-10mu\lambda}} 

\let\<=\langle
\let\>=\rangle

\let\+=\uparrow

\begin{document}   
\title{Twin Higgs WIMP Dark Matter} 
\author{Isabel Garc\'ia Garc\'ia}
\email{isabel.garciagarcia@physics.ox.ac.uk}
\affiliation{Rudolf Peierls Centre for Theoretical Physics, University of Oxford,
1 Keble Rd., Oxford OX1 3NP, UK}
\author{Robert Lasenby}
\email{robert.lasenby@physics.ox.ac.uk}
\affiliation{Rudolf Peierls Centre for Theoretical Physics, University of Oxford,
1 Keble Rd., Oxford OX1 3NP, UK}
\author{John March-Russell}  
\email{jmr@thphys.ox.ac.uk}  
\affiliation{Rudolf Peierls Centre for Theoretical Physics, University of Oxford,
1 Keble Rd., Oxford OX1 3NP, UK}
\affiliation{Department of Physics, Stanford University, Stanford, CA 94305, USA}

\begin{abstract}
Dark matter (DM) without a matter asymmetry is studied in the context of Twin Higgs (TH) theories in which the LHC naturalness problem is addressed.  These possess a twin sector related to the Standard Model (SM) by a (broken) $\mathbb{Z}_2$ symmetry, and interacting with the SM via a specific Higgs portal.  We focus on the minimal realisation of the TH mechanism, the Fraternal Twin Higgs, with only a single generation of twin quarks and leptons, and $SU(3)'\times SU(2)'$ gauge group.   We show that a variety of natural twin-WIMP DM candidates are present (directly linked to the weak scale by naturalness), the simplest and most attractive being the $\tau^\prime$ lepton with a mass $m_{\tau^\prime} > m_{\rm Higgs}/2$, although spin-1 $W^{\prime\pm}$ DM and multicomponent DM are also possible (twin baryons are strongly disfavoured by tuning).  We consider in detail the dynamics of the possibly (meta)stable glueballs in the twin sector, the nature of the twin QCD phase transition, and possible new contributions to the number of relativistic degrees of freedom $\Delta N_{\rm eff}$.  Direct detection signals are below current bounds but accessible in near future experiments.  Indirect detection phenomenology is rich and requires detailed studies of twin hadronization and fragmentation to twin glueballs and quarkonia and their subsequent decay to SM, and possible light twin sector states.
\end{abstract}  
\maketitle 
\section{\label{intro}Introduction}

Models based on the Twin Higgs (TH) mechanism
\cite{TwinHiggs}\cite{Chacko2005,ChackoNomura,Barbieri2005,Craig:2013fga} address
the LHC fine-tuning problem and solve the little hierarchy problem by
introducing a \emph{twin} sector, that is, in its simplest realisation,
a copy of the Standard Model (SM), regarding both field content and
interactions. At tree level, the Higgs sector of the theory respects
a global $SU(4)$ symmetry (in fact an $O(8)$ symmetry in the most
attractive cases \cite{Chacko2005}) acting on the components of the
pair of Higgs doublets $(H,H^\prime)$, where $H^\prime$ is the
twin Higgs doublet (throughout primes denote objects in the twin
sector). A discrete $\mathbb{Z}_2$ between the two sectors ensures
equality of their couplings, which results in $SU(4)$-symmetric
radiative corrections to the Higgs mass squared. The $SU(4)$ symmetry
is broken at one loop order by radiative corrections to the Higgs
\emph{quartic} coupling and the SM Higgs is realised as a naturally
light pseudo-Nambu-Goldstone boson of the approximate $SU(4)$. The
$\mathbb{Z}_2$ symmetry needs to be broken, explicitly or otherwise,
for the SM Higgs to acquire a phenomenologically viable vacuum
expectation value (vev), for an exact $\mathbb{Z}_2$ would imply that
the vev's in the two sectors are equal, a possibility that is excluded
by Higgs-coupling measurements. Denoting the SM and TH vev's as $v
\approx 246 \ \gev$ and $f$ respectively, the fine-tuning arising from
this difference of vev's is $\sim 2 v^2 / f^2$, i.e.\ a mild $\sim 20
\%$ tuning for the minimum experimentally allowed ratio $f/v \approx
3$. The physical light Higgs state, $h$, is shared between the SM and
twin sectors with couplings to SM and twin sector states
modified by $\cos(v/f) \simeq 1- v^2/2f^2$ and $\sin(v/f) \simeq v/f$
respectively. It is important to also bear in mind that the TH theory
needs to be UV completed at some cutoff scale $M_{\rm UV} \leq 4 \pi f$
(for definiteness we take $M_{\rm UV} \approx 5\tev$).

For the TH mechanism to operate, the twin sector does not need to
be an exact copy of the SM, a reduced field content sufficing. This
simplified version -- the Fraternal Twin Higgs (FTH) \cite{FraternalTH}
-- has as its minimal ingredients twin weak $SU(2)^\prime$ and colour
$SU(3)^\prime$ interactions, and a twin third generation consisting
of top and bottom quarks $Q^\prime$, $t_R^\prime$ and $b^\prime_R$,
a lepton doublet $L^\prime$ required by $SU(2)^\prime$ anomaly
cancellation, and a twin Higgs doublet $H^\prime$. Right handed twin
leptons may be added to the theory rendering the leptons massive,
although they are not required by the TH mechanism. For the TH
mechanism to still be effective without introducing significant extra
tuning, the twin top Yukawa $y_{t^\prime}$ can only differ by at most
$1 \%$ from $y_t$ for $M_{\rm UV} \approx 5\tev$. The need for gauged
twin $SU(3)^\prime$ then becomes apparent: radiative corrections to
$y_t$ and $y_{t^\prime}$ could make them differ significantly at the
weak scale even if they coincided at $M_{\rm UV}$. A gauge coupling
$g_3^\prime$ differing by less than $\sim 15\%$ from $g_3$ at the cutoff
\cite{FraternalTH} ensures that the running of $y_t$ and $y_{t^\prime}$
is close enough. However, the running of the $g_3^\prime$ and $g_3$
couplings, and thus the dynamical scales, differ because of the
different field content and masses of the two sectors. With only one
quark generation in the twin sector, and for the allowed range of values
of $g_3^\prime$, we have a twin confinement scale
$\Lambda_{\rm QCD}^\prime \sim 0.5 - 20\gev$
\cite{FraternalTH}. (Unless otherwise stated we will take our default
value as $\Lambda_{\rm QCD}^\prime \approx 3\gev$; the preference for
this choice, or larger, is justified in Section~VIII.) The presence
of an $SU(2)^\prime$ gauge group in the twin sector also follows
from fine-tuning considerations, with the twin $g_2^\prime$ coupling
allowed to differ by at most $\sim 10\%$ from the SM value. A gauged
$U(1)^\prime$ is \emph{not} necessary from a naturalness perspective,
although it remains an accidental global symmetry of the twin sector.

In this paper, we explore the possibilities for dark matter (DM) in
TH scenarios, focusing on the minimal FTH models in the case where a
matter-antimatter asymmetry in the twin sector is \emph{not} present and
where $U(1)^\prime$ is a global symmetry. (We reserve the
study of both twin \emph{asymmetric} DM, and the effects of a gauged
$U(1)^\prime$, for a companion paper \cite{AsymmetricTwinDM}.) Crucially
for our later discussion, intrinsic to the success of the TH mechanism
is the fact that the twin sector interacts with the SM {\it via the
Higgs portal with a strength determined by} $f/v$, leading to specific
predictions for DM signals. In addition, we will show that the most
attractive twin DM candidates in the absence of an asymmetry are the
twin leptons, which naturally have a `twin-WIMP miracle' as they freeze
out via twin weak interactions whose strength is set by $g_2^\prime
\simeq g_2$ and $G_F^\prime = (v/f)^2 G_F$ both of which are directly
tied to SM weak interaction values by naturalness.
Since the Higgs portal forces the SM and twin sector to be in 
equilibrium until temperatures well below the EFT cutoff,
we have a purely thermal freeze-out scenario, at least
for states that are not UV relics.

As asymmetric DM is not our concern here, we will primarily focus on the heavy
twin quark limit, i.e. $m_{t^\prime}, m_{b^\prime} \gg \Lambda_{\rm
QCD}^\prime$, which arises naturally for $y_{t^\prime} = y_t$ and
$y_{b^\prime} \approx y_b$ and for the values of $f/v\geq 3$ that are
allowed. Collider signals of the TH model in this regime were treated in detail
in~\cite{FraternalTH}, and also considered in~\cite{Curtin:2015fna}.

Finally we remark that the idea of a mirror world, either exact or
partial, has a long and rich history~\cite{Blinnikov:1982eh,Carlson:1987si,Foot:1991bp}, see e.g. \cite{Okun:2006eb}.
(For a recent review of aspects of mirror world physics we refer the
reader to \cite{Foot:2014mia}.) Often such theories lead to a variety
of interesting DM candidates with overlap with those studied here.
Here we are considering a particular, approximate mirror world interacting
with the SM via the Higgs portal, as directly motivated by the LHC
naturalness problem.
Previous investigations of symmetric dark matter candidates in models
with a similar philosophy include~\cite{Poland:2008ev} and~\cite{Hedri:2013ina}.

\section{\label{sec:stablestates} Stable  \& Metastable Twins}

At temperatures well below the $SU(2)'$ phase transition, where anomaly
effects are exponentially small, the FTH model has an accidental $U(1)$
global symmetry associated to twin baryon number B$^\prime$.
If Majorana mass terms for the right-handed twin leptons
are forbidden (for example, due to a discrete symmetry), then there are
also accidental $U(1)$ twin lepton number and twin `charge' symmetries, with
associated conserved numbers L$^\prime$ and Q$^\prime$.
Ultimately, we might expect these global symmetries to be
explicitly broken by higher dimensional operators, possibly connected
with Planck-scale physics, or by terms from the UV completion of the
TH models that connect the SM and twin sectors in ways beyond the
TH-mandated Higgs portal interaction itself.
For the purposes of this
work we assume that these new interactions are sufficiently weak that the
lightest states carrying B$^\prime$, L$^\prime$ and Q$^\prime$ are stable on
timescales $\gsim 10^8 H_0^{-1}$, although decaying DM is a natural
possibility in TH models.

The discrete symmetries $P$ and $C$ in the
twin sector are maximally violated by $SU(2)^\prime$ interactions
but, in principle, $CP$ can remain conserved.  Although we focus on the
$CP$-preserving scenario, breaking of $CP$ in the twin sector, due,
say, to an un-cancelled $\bar{\theta}_{\rm QCD'}$ term, is allowed and
can have important consequences.  We mention the changes this makes
where appropriate.

We first consider the case where $m_{\nu^\prime} \leq m_{\tau^\prime}$,
$m_{\nu^\prime}, m_{\tau^\prime} < m_{W^\prime}$, implying that both
$\nu^\prime$ and $\tau^\prime$ are stable states and therefore automatic
DM candidates. We will focus on the regime of heavy $\tau^\prime$ ($
m_{\tau^\prime} \gsim m_h/2$), whereas we will allow $\nu^\prime$ to be
heavy or effectively massless. When $\nu^\prime$ is effectively massless
it will behave as dark radiation (DR), contributing to the number of
effective neutrinos, $N_{\rm eff}$. (We discuss the issue of $N_{\rm
eff}$ in detail in Section~VIII.) If $m_{\nu^\prime} + m_{\tau^\prime}
> m_{W^\prime}$, the twin $W^{\prime \pm}$ gauge bosons will also be
stable, as automatically $m_{t^\prime} + m_{b^\prime} > m_{W^\prime}$ in
TH models, and therefore $W^{\prime \pm}$ could contribute significantly
to the DM density.

In the strong sector, things are more involved. One stable state is the
lightest twin baryon, $\Delta^\prime$, made out of three $b^\prime$
quarks in a spin 3/2 state.
In the absence of an asymmetry in the twin
sector, $b^\prime \bar{b^\prime}$ pairs annihilate efficiently into
gluons rendering $\Delta^\prime$ irrelevant as a
DM candidate unless $m_{b^\prime}\gsim 1\tev$, a case we discuss in
Section~V (if the freeze-out temperature of the $b^\prime$ quark-antiquark annihilations is below the
phase transition temperature then $\Delta^\prime$ (anti)baryons annihilate efficiently into
glueballs and quarkonia).  
However, twin QCD glueball states are themselves of potential
interest depending upon the UV completion of the TH theory. In the heavy
quark regime the spectra of glueball and quarkonia are well separated
(and relatively well known). The lightest glueball, whose mass is $m_0
\approx 6.8 \Lambda_{\rm QCD}^\prime$ \cite{Morningstar,Chen}, has
$J^{PC}$ quantum numbers $0^{++}$ and therefore mixes with the
Higgs, with mixing angle \cite{FraternalTH}
\begin{equation}
	\theta = \frac{\alpha_3^\prime v \mathcal{F}_0}{6 \pi f^2 (m_h^2 - m_0^2)} \approx \frac{v m_0^3}{8 \pi^2 f^2 m_h^2}
\label{eq:thetah}
\end{equation}
with $\mathcal{F}_0$ the $0^{++}$ glueball decay constant given by $\mathcal{F}_0 = 3.06 m_0^3 / (4 \pi \alpha^\prime_3)$ \cite{Chen}
and we have assumed $m_h^2 \gg m_0^2$ in the final step.
It therefore promptly decays to light SM states ($\tau_{0^{++}} \sim 4\times10^{-10} \ {\rm s}$ if we take $\Lambda_{\rm QCD}^\prime= 3\gev$ and $f/v=3$).  In the case where the twin neutrino is light compared to $\Lambda_{\rm QCD}^\prime$, all other glueballs decay, via $SU(2)^\prime$ and heavy quark-loop induced interactions, to (eventually) combinations of the $0^{++}$ glueballs and $\bar {\nu'} \nu^\prime$ in appropriate angular momentum states, so leaving no stable twin-QCD states apart from $\Delta^\prime$.

In the case where both $\tau^\prime$ and $\nu^\prime$ are heavy,
such that twin-lepton pairs are not kinematically accessible in glueball decays, another
two glueballs become potentially relevant:
a $0^{-+}$ glueball, with mass $m_{0^{-+}} \approx 1.5 m_0$ \cite{Morningstar,Chen},
and a $1^{+-}$ glueball, with mass $m_{1^{+-}} \approx 1.7 m_0$~\cite{Morningstar, Chen}.
These glueballs can be potentially long-lived metastable states, possibly with cosmologically
long lifetimes, though this sensitively depends on the issues of additional SM--twin-sector interactions
arising from the UV completion and/or twin $CP$-violation.  
As we discuss in Section~VI the freeze-out abundance of these
glueballs is very small, $\sim 10^{-8} \rho_{\rm DM}$, so they
are not significant gravitationally, though they may be constrained by
CMB or cosmic ray observations if their lifetimes are long.
Also, as we discuss in Section~VII, both $0^{-+}$ and $1^{+-}$ glueballs have
the potential to produce novel indirect detection signals if they decay
with lifetimes in the ranges $\tau \sim 1- 10^{3} \ {\rm s}$ or $\tau \sim
10^{11} - 10^{12} \ {\rm s}$.  Whether such lifetimes are achieved
depends on the value of $\Lambda_{\rm QCD}^\prime$, and on the UV
completion, a subject we reserve for Section~VI.

\section{ Twin $SU(3)$ phase transition}

Before we can calculate the relic densities of the stable twin states, we
first must consider the nature of the twin QCD phase transition, and whether
it leads to significant dilution of relics by entropy production.

If $m_{b'} \lesssim 8 \Lambda_{\rm QCD}'$, then the
one-dynamical-quark-flavour $\rm QCD'$ phase transition is a smooth
cross-over~\cite{Pisarski:1983ms,Alexandrou:1998wv,Fromm:2011qi}, with no significant
non-equilibrium dynamics.  As $m_{b'}$ is increased, the transition
becomes second order at a critical value $\sim 8 \Lambda_{\rm QCD'}$
(it should be noted that this upper limit is potentially
uncertain by a few times $\Lambda_{\rm QCD}'$),
and above that is (weakly) first order, as demonstrated
by analytical arguments~\cite{Pisarski:1983ms,Yaffe:1982qf}
and lattice studies~\cite{Panero:2009tv}.

Investigating the dynamics of the first-order phase transition in 
the heavy-quark case, lattice studies give a critical temperature of $T_c \simeq 1.26
\Lambda_{\rm QCD '}^{\bar{\rm MS}}$~\cite{Borsanyi:2012ve}, bubble wall
surface tension $\sigma \simeq 0.0155 T_c^3$, and latent heat 
$\rho_L \simeq 1.4 T_c^4$~\cite{Beinlich:1996xg}.
Since the pressure is continuous across the phase transition, the
pressures of the confined and unconfined phases are equal at $T_c$.
If the unconfined phase manages to supercool to a temperature
$T = (1 - \delta) T_c$, then a bubble of confined phase can grow
due to the pressure difference between the phases, if it is large enough
to overcome the surface tension, i.e.\ if it has radius $R \ge R_c = \frac{2 \sigma}{\Delta P}$.
The free energy cost of a `critical bubble' of radius $R_c$ is
\begin{equation}
\Delta F_c = 4 \pi R_c^2 \sigma - \frac{4}{3} \pi R_c^3 \Delta P
= \frac{16 \pi}{3} \frac{\sigma^3}{(\Delta P)^2}~.
\end{equation}
Assuming that the supercooling is small, $\delta \ll 1$, so that
\begin{equation}
\Delta P = P_G(T) - P_g(T) \simeq - \delta T_c (s_G(T_c) - s_g(T_c))
= \delta \, \rho_L~,
\end{equation}
where $P_G$ is the pressure in the confined phase, $P_g$
is the pressure in the unconfined phase, and $s_G, s_g$
are the corresponding entropy densities,
we have 
\begin{equation}
\frac{\Delta F_c}{T} \simeq \frac{16 \pi}{3} 
\frac{\sigma^3}{\rho_L^2 T_c} \delta^{-2}
\simeq 3 \times 10^{-5} \delta^{-2}~.
\end{equation}
The rate per unit volume of bubble nucleation due
to thermal fluctuations is set by $\Gamma/V \sim T^4 e^{-\Delta F_c / T}$ ~\cite{Csernai:1992tj}.
This becomes comparable to $H^4$,
i.e.\ one event per Hubble volume per Hubble time, when
$\Delta F_c /T \lesssim \log\left(T^4/H^4\right)$. Assuming that $\delta \ll 1$,
this is equivalent to $3 \times 10^{-5} \delta^{-2} \lesssim 4 \log (T_c/H(T_c)) 
\simeq 160$, so $\delta \gtrsim \delta_n \equiv 4 \times 10^{-4}$. Since $e^{- \Delta F_c / T}$
grows by an $e$-fold with a $\sim 10^{-6}$ drop in $\log T$,
the nucleation rate quickly becomes large as $T$ drops.

This extremely large nucleation rate means that, once $\delta$ is somewhat larger
than $\delta_n$, the latent heat released from bubble nucleation and
expansion must heat us back above the nucleation temperature, or the
transition must complete entirely in a very small fraction of a Hubble
time (in the most extreme limit, for $\delta \sim 10 \delta_n$ we have $e^{-\Delta F_c/T} \sim
\mathcal{O}(1)$, and nucleation effectively happens everywhere at once). 
The rapid-completion case would arise if e.g.\ the heat capacity
of the supercooled unconfined phase was very high, so that the expansion
necessary to cool it from $T_c$ to $(1 - \delta) T_c$ removed
almost all of the latent heat of the transition.
In the former case, the confined phase bubbles grow, releasing
latent heat, until the temperature has been brought back up
to $T_c$, where the pressures are equal. Once there,
the confined phase bubbles can only grow as Hubble expansion
removes energy from the mixture. In this way they grow until they occupy the entire
volume, with the temperature remaining constant at $T_c$.

In either case, the out-of-equilibrium part of the phase transition,
during which there is a pressure difference driving confined phase
bubble expansion, occurs at temperatures only very slightly below $T_c$,
so can only produce a small amount of entropy. The maximum possible
increase in the overall entropy density occurs in the rapid-transition
case, where the pressure-driven growth transforms all of the volume from the
unconfined to the confined phase, in which case the overall increase in
entropy density is
\begin{equation}
\Delta s \simeq \frac{\Delta P}{T_c} \simeq  {\rm few}  \times \delta_n \frac{\rho_L}{T_c}
\sim 10^{-3} T_c^3~.
\end{equation}
In the (naively more likely) case where most of the transition occurs
in quasi-equilibrium, the maximum entropy production is even smaller,
$\Delta s \propto \delta_n^2 \rho_L/T_c \sim 2 \times 10^{-7} T_c^3$,
since only a fraction $\propto \delta_n$ of the volume is converted by
out of equilibrium growth.
Either way, there is no significant effect on relic densities.
If the transition is simply a cross-over, as occurs for light enough
$b'$, then there is again no significant entropy production.
For a crossover or a quasi-equilibrium transition,
there will not be significant gravitational radiation
production, rendering the FTH model less encouraging
for gravity wave signals than the other scenarios
discussed in~\cite{Schwaller:2015tja}.

We also remark that if twin-$CP$ is violated by $\theta_{\rm QCD'}\neq 0$
in the pure $SU(3)$ case with no light quarks, the transition remains
first order but with parameters such as the critical temperature,
$T_c(\theta)$, and the latent heat, now depending on $\theta_{\rm QCD'}$ in a
periodic fashion but otherwise poorly constrained, either analytically
\cite{Anber:2013sga} or from lattice studies \cite{D'Elia:2012vv}.
These studies indicate that for $\theta_{\rm QCD'}\ll \pi$ the
transition temperature decreases $T_c(\theta)<T_c(0)$, while the
strength of the transition slowly increases. However the properties
of the phase transition at $\theta_{\rm QCD'}\neq 0$ are not firmly
established, especially once $\theta_{\rm QCD'}$ is not small. Thus
to be conservative, in the cases with twin-$CP$ violation we take
$\theta_{\rm QCD'}\ll \pi$, a limit that is sufficient for our purposes.

\section{\label{sec:tauDM} Twin Tau Dark Matter}

Having argued in Section III that the twin-QCD phase transition
leads to no significant dilution of relics by entropy production
we now proceed to calculate the freeze-out density of the stable
twin-sector states. We start with the simplest case, that of the
twin $\tau$ lepton (since $U(1)^\prime$ is not gauged, the situation
for $\nu '$ is identical to that of $\tau '$ assuming a suitable
Yuakawa coupling giving a Dirac mass). In most of parameter space the
annihilation of $\tau '$'s dominantly proceeds via twin-$SU(2)$
weak interactions into the (assumed lighter)
$b^\prime$-quarks/quarkonia and $\bar {\nu^\prime} \nu^\prime$
pairs. Annihilation via the Higgs, with couplings that are given by
$\frac{y_{\tau}^\prime}{\sqrt{2}}\frac{v}{f} h \bar{\tau '} {\tau
'}$, is subdominant apart from a narrow resonance region around
$m_{\tau^\prime} \sim m_h / 2$.

\begin{figure}[h!]
  \includegraphics[scale=0.53]{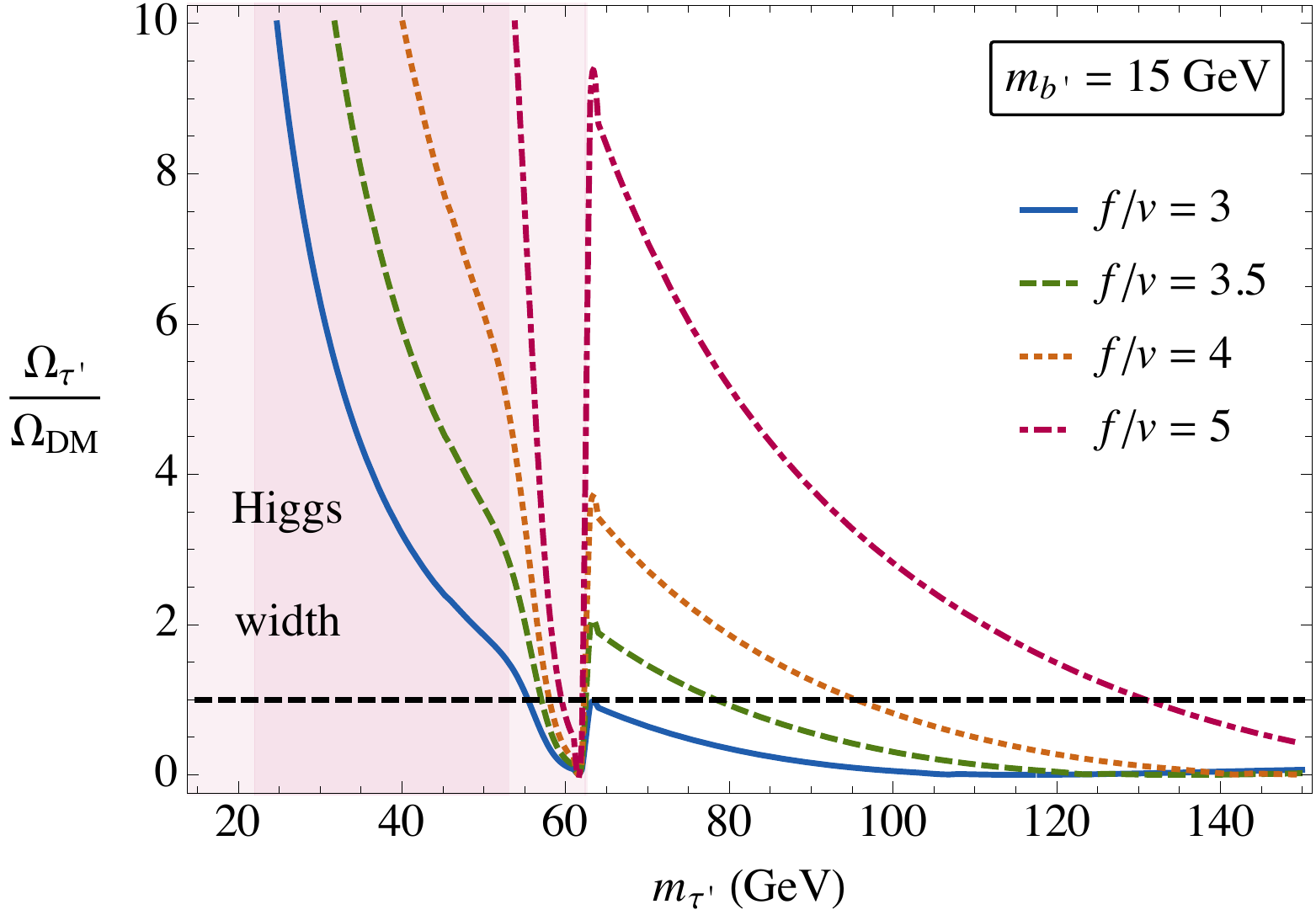}
	\caption{\label{fig:tauDM} Contribution to the energy density of the Universe from $\tau^\prime$ species normalized to the observed DM
	energy density as a function of $m_{\tau^\prime}$ for different values of $f/v$.
	Light (dark) pink area indicates the 2-sigma bounds from invisible Higgs width and modified couplings to visible sector particles in the case $f/v = 3$ ($3.5$),
	whereas $f/v=4,5$ remain unconstrained in the region of parameter space shown. 
Note that, if $\Lambda_{\rm QCD}'$ is large enough so that $m_0 \gtrsim
2 m_{b^\prime}$, then annihilations of low-mass $\tau'$s will have
significant non-perturbative corrections. However, this regime generically 
leads gives too high a $\tau'$ density, so is not of primary concern here.}
\end{figure}

Figure~{\ref{fig:tauDM}} shows the contribution to the present energy
density of the Universe from $\tau^\prime$ species, normalized to the
observed DM density for different values of $f/v$. 
This calculation, along with the other relic
density calculations throughout this paper,
was performed using the MicrOMEGAs software package~\cite{Belanger:2013oya}. For $f/v \approx
3$, the least tuned case, the observed DM density is obtained for
$m_{\tau^\prime} \approx 63 \gev$.
Larger values of the ratio $f/v$, which imply worse tuning, result in
larger DM masses.\footnote{
For concreteness, we take $m_{b^\prime} \approx 15 \gev$,
which saturates the experimental
bound coming from constraints on the Higgs width for $f/v \approx 3$.
This also corresponds to the $Z_2$ symmetric value $(f/v)m_b$ for $f/v \approx 3$.
The twin bottom Yukawa in the FTH is only constrained by tuning
to be $\ll y_{t'}$. Also, as long as $m_{b'}, \Lambda_{\rm QCD}' \ll m_{\tau'}$,
the $\tau'$ relic density is insensitive to the precise $b'$ mass.
Thus,
different (but still sufficiently light) values of $m_{b'}$ would not
affect our conclusions.
}


We emphasise that, for (symmetric) twin DM candidates with
$\mathcal{O}(0.1)$ Yukawa couplings, we can obtain the correct relic density
since the couplings and mass of the $W^\prime$ bosons are
set by the TH mechanism to be $g_2^\prime \simeq g_2$ and $M_{W^\prime}
\simeq (f/v) M_{W^\pm}$. These are constrained by naturalness to be
close to SM weak interaction values,  
giving rise to a natural `twin-WIMP-miracle'.

\begin{figure}[h!]
  \includegraphics[scale=0.55]{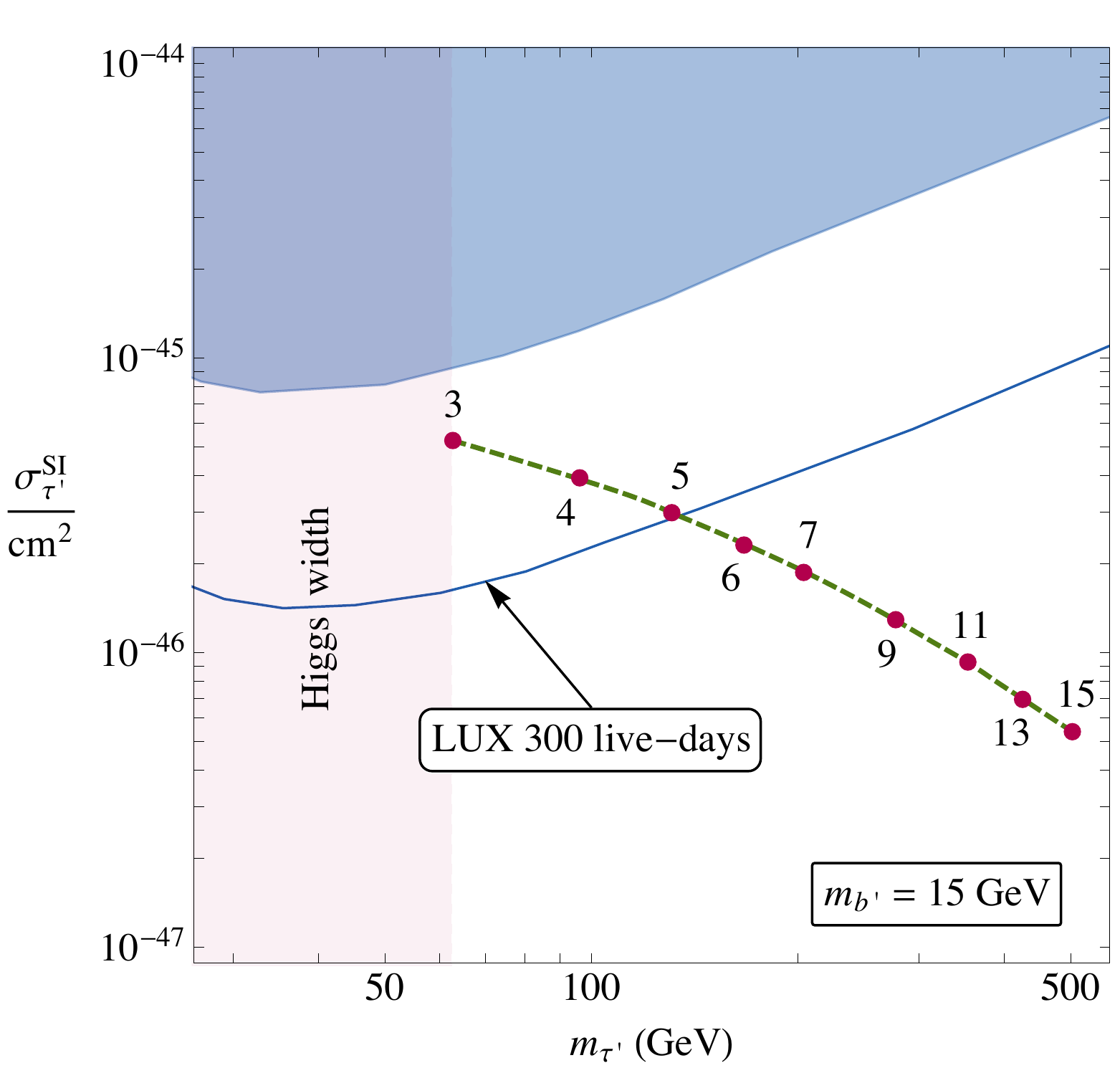}
  \caption{\label{fig:tauDMDD}
	Dashed green line is the spin independent scattering cross section per nucleon for $\tau^\prime$ DM for those
	values of $m_{\tau^\prime}$ and $f/v$ such that the correct DM density is obtained.
	Red dots point out particular values of $f/v$ (indicated in numbers).
	Blue area: LUX current bounds \cite{LUXresults}; blue line: LUX projected sensitivity (300 live-days) \cite{LUXprojection};
	pink area: region of parameter space ruled out by bounds on the invisible
Higgs width, and on modified couplings to visible sector particles~\cite{FraternalTH}.}
\end{figure}

Turning now to direct detection, scattering of $\tau'$ with SM nuclei
occurs, at tree level, via Higgs exchange, and this process sets
the scattering cross section in direct detection experiments. In
Figure~\ref{fig:tauDMDD} we show the spin independent scattering cross
section per nucleon off SM nuclei for $\tau^\prime$ DM as a function
of $m_{\tau^\prime}$ and for values of $f/v$ such that the correct DM
abundance is obtained. For $f/v \approx 3 - 5$ (tuning $20 - 8 \%$), the
predicted direct detection signatures fall below current experimental
bounds \cite{LUXresults} but above the region of parameter space that
will be probed in the very near future by LUX \cite{LUXprojection}.
Larger (more tuned) values of $f/v$ will be probed by next-generation
experiments such as LUX-ZEPLIN (LZ) \cite{LZ}.

\section{\label{sec:multicomponentDM} Multicomponent, $W^\prime$, \& $\Delta^\prime$ Dark Matter}

In the case where the sum of $\tau^\prime$ and $\nu^\prime$ masses is larger than the $W^\prime$ mass, the latter is not able to
decay. In the regime where $m_{\tau^\prime} \sim m_{\nu^\prime}$ and $m_{\tau^\prime}, m_{\nu^\prime} < m_{W^\prime}$, this implies that
all three states are stable and may significantly contribute to the DM energy density, opening a possibility for a 3-component DM scenario.

\begin{figure}[h!]
  \includegraphics[scale=0.52]{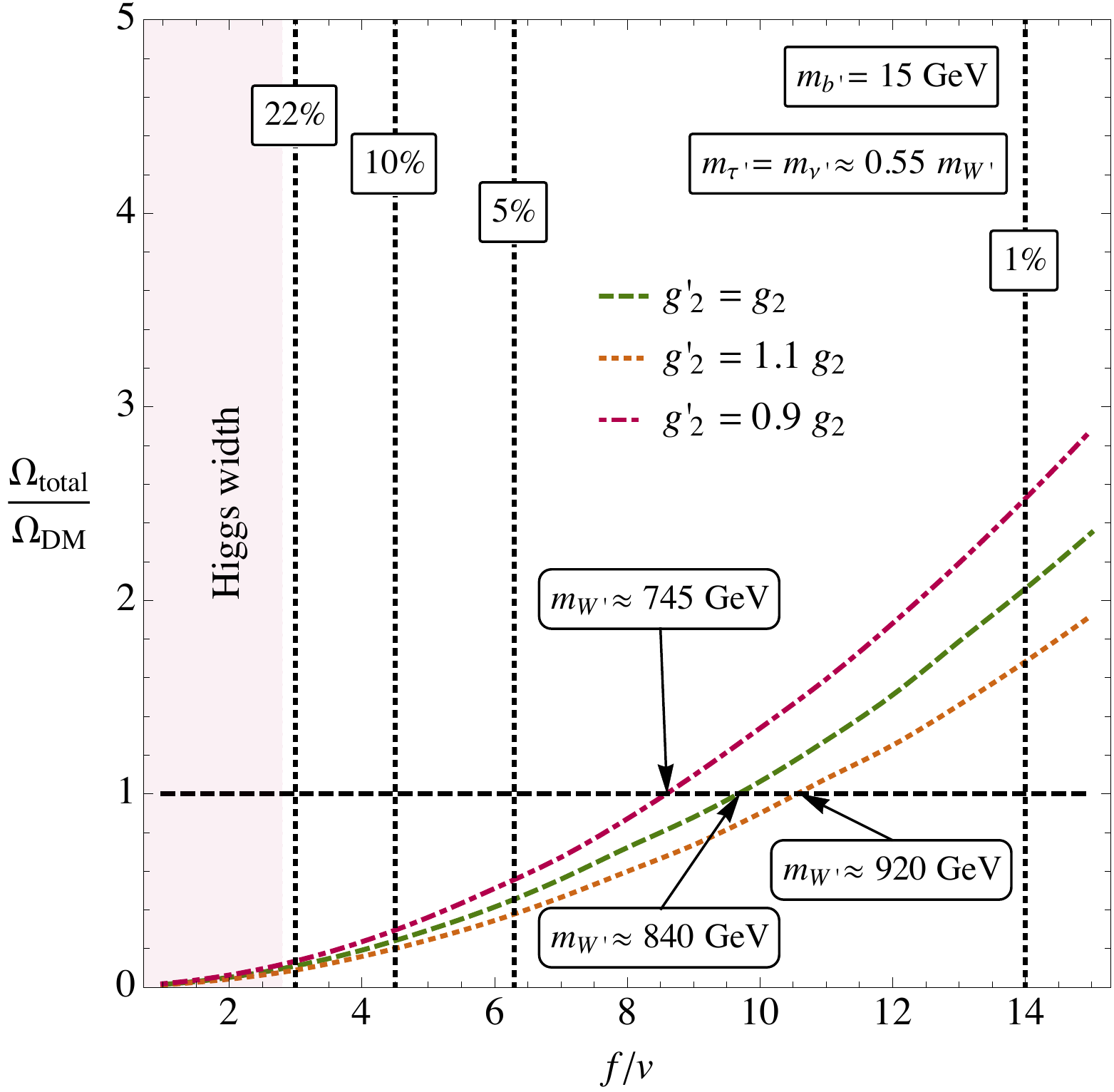}
\caption{\label{fig:multicomponentDM}
		Contribution to the energy density of the Universe from $\tau^\prime$, $\nu^\prime$ and $W^\prime$, species normalized to the observed DM
		energy density, as a function of $f/v$ and for different values of the twin weak coupling $g_2^\prime$.
		Vertical lines represent tuning contours.
		We indicate the $W^\prime$ mass for the three different values of $g_2^\prime$
		considered when the right DM density is achieved.} 
\end{figure}

Figure~\ref{fig:multicomponentDM} shows the contribution to the DM
energy density of these three particle species (normalized to the
observed value) for different values of the twin weak
coupling, that we allow to vary by $10 \%$ from its central
value $g_2^\prime = g_2 \approx 0.64$. For concreteness, we have taken
$m_{\tau^\prime} = m_{\nu^\prime} \approx 0.55 \ m_{W^\prime}$, with
$m_{W^\prime} = g_2^\prime f / 2$. As one can see, the observed DM
energy density is only achieved for relatively large values of the ratio
$f/v$, where the fine-tuning is in the range $5 \%$ to $1 \%$.
This occurs since $\tau'$ and $\nu'$ are forced to be heavier than considered
in the $f/v \simeq 3$ case, so
their annihilation cross sections set by $m^2/f^4$ are larger ---
to compensate, $f$ must be increased.

As can be read off from Figure~\ref{fig:multicomponentDM}, the correct DM abundance is obtained for $f/v \approx 9.7$ for $g_2^\prime = g_2$, which implies a tuning of approximately $2 \%$.  In this case, $\tau^\prime$ and $\nu^\prime$ species would contribute roughly $75 \%$ to the DM energy density, with $W^\prime$ species making for the remaining $25 \%$.

Regarding direct detection experiments, the predicted spin independent scattering cross section per nucleon for all three particle species 
is of order $\sim 10^{-46} {\rm cm}^2$ for values of the masses that result in the correct DM abundance.
This lies around an order of magnitude below LUX current projected sensitivity for the range of masses considered, which means that next-generation direct detection experiments such as LZ \cite{LZ} will be able to cover the relevant region of parameter space.

Small variations of the values of $\tau^\prime$ and $\nu^\prime$ masses around the case we have considered do not make a significant
difference to our conclusions, except when $m_{\tau^\prime} \sim m_{\nu^\prime}$ but $m_{\tau^\prime} + m_{\nu^\prime} < m_{W^\prime}$.
In this case, the $W^\prime$ is no longer stable and then only $\tau^\prime$ and $\nu^\prime$ species would contribute to the DM density.
In this 2-component DM scenario, sufficient annihilation requires $m_{\tau^\prime}$ and $m_{\nu^\prime}$ to be in the mass range above $m_h/2$,
automatically evading invisible Higgs width constraints.
The different contribution to the DM density from the two particle species would depend solely on the ratio of their masses: for equal masses, both
components would contribute $50 \%$, whereas if they differ by approximately $10 \gev$ the right DM abundance requires $m_{\nu^\prime} \approx 70 \gev$
(therefore $m_{\tau^\prime} \approx 80 \gev$) and $\nu^\prime$ and $\tau^\prime$ species would make for $65 \%$ and $35 \%$ of DM respectively.
Regarding direct detection signals, this 2-component scenario is analogous to the single-component case discussed in Section IV,
for interactions between the DM species and the visible sector proceed only via Higgs exchange.

Finally we turn to the most complicated of the possible twin
DM candidates, the $\Delta^\prime$ baryon. Although for light
$b^\prime$ quarks, and in the absence of a matter-antimatter asymmetry
(a subject of a companion paper \cite{AsymmetricTwinDM}), the spin-3/2
$\Delta'$ baryons efficiently annihilate to glueballs and quarkonia, leaving an
uninterestingly small freeze-out density, this is no longer the case
if the ${b^\prime}$-quarks, and thus the $\Delta^\prime$ baryons, are
sufficiently heavy, $m_{b^\prime} \gsim 1 \tev \gg \Lambda^\prime_{\rm
QCD}$.

To estimate the freeze-out density of such states let us consider the
case where the freeze-out temperature is well above $\Lambda^\prime_{\rm
QCD}$, a situation that applies if $m_{b^\prime}$ is sufficiently large.
In this case we may self-consistently work with $b^\prime$ quarks and
twin-gluons, and first calculate the freeze-out density of $b^\prime$
quarks, via a leading annihilation cross section to two gluons that
parametrically goes as $\sigma v \sim (\alpha^\prime_3/m_{b^\prime})^2$.
We find that a numerical evaluation of the annihilation
rate leads to a freeze-out temperature $T_f \sim m_{b^\prime} /30$
and gives a substantial freeze-out density of $b^\prime$ quarks and
anti-quarks only once $m_{b^\prime}\gsim 1\tev$, which implies very
large $f/v\gsim 30$ and thus a very badly tuned TH model (here we have
taken $y_{b^\prime} \lsim 0.2$ so as not to have yet further tuning at
1-loop). We therefore come to the conclusion that the $\Delta^\prime$
baryon in the absence of a matter-antimatter asymmetry is a poor
DM candidate in TH models. (We remark that the freeze-out density
estimated this way provides the most optimistic estimate of the minimum
$b^\prime$ mass.  The reason is that the $b^\prime$ quark and anti-quark
densities do not simply translate, via a factor of $1/3$, into the final
freeze-out density of $\Delta^\prime$ baryons and anti-baryons. Post
twin-confinement only a proportion of $b^\prime$-(anti-)quarks end up
in $\Delta^\prime$ (anti-)baryons, and thus are stable relics, compared
to the number that form $b' \bar {b '}$ quarkonia and therefore quickly
decay. We expect this proportion to be an ${\cal O}(1)$ number but we
are not aware of any reliable calculation of the ratio.)

\section{\label{sec:glueballrelic}  Glueball Metastability and Relic Density}

If the twin QCD sector had no couplings to other states,
then after the phase transition, the glueball bath would
behave roughly as non-relativistic strongly-self-interacting DM
(since all of the glueballs have mass $\gtrsim 5.6 T_c$). That is,
while number-changing (e.g.\ $3 \rightarrow 2$) interactions were still fast enough
to maintain number equilibrium, its temperature
would decrease only logarithmically with the scale factor~\cite{Carlson:1992fn},
with its energy density decreasing as $\rho \propto 1/(a^3 \log(a/a_0))$,
where $a_0$ is a constant.

However, even in the absence of light twin sector states, the Higgs
mixing portal with the SM necessarily provides a coupling to lighter
states, and in particular, means that most of the glueballs decay
rapidly to the SM.  The question then becomes whether the relic population
of (meta)stable glueballs is small enough not to be cosmologically dangerous.

To calculate the relic population of (meta)stable glueballs, 
we need to know the point at which their number-changing interactions
freeze out. The last such interactions
to freeze out will be processes with two particles in the initial state.
Inspecting the glueball spectrum, if states have their equilibrium abundances
then we expect the fastest such processes to be annihilation
to lighter glueballs, e.g. $0^{-+}0^{-+} \rightarrow 0^{++}0^{++}$.
We can therefore perform the usual calculation for the relic density
of annihilating DM. If we parameterise the annihilation cross
section as $\langle \sigma v \rangle \sim C / (\Lambda_{\rm QCD}')^2$, then
the present-day relic density is (assuming no significant
entropy injection)
\begin{equation}
\Omega h^2 \simeq 3.8 \times 10^{-10} (\Lambda_{\rm QCD}^\prime / \GeV)^{2} C^{-1}~,
\end{equation}
in comparison to the present-day DM relic density $\Omega_{\rm DM} h^2 \simeq 0.12$~\cite{Ade:2015xua}. This assumed that the SM and glueball temperatures
were equal at the freeze-out time. This will be the case when the decay
rates of some of the glueballs are sufficiently fast---roughly,
faster than the Hubble rate at freeze-out. If it is not the case, then as described above,
the glueballs will be at a higher temperature than the SM, as their
temperature will have fallen only logarithmically with the scale factor.
Comparing the $0^{++} \rightarrow {\rm SM}$ decay rate (from the mixing
of Eq(\ref{eq:thetah}))
to the Hubble rate at freeze-out, the former is larger for $\Lambda^\prime_{\rm QCD} \gtrsim 0.6 \GeV$,
so we are in the fast-decay regime for most of the $\Lambda^\prime_{\rm QCD}$ range of interest.

As $\Omega / \Omega_{\rm
DM} \simeq {\rm few} \times 10^{-9} (\Lambda^\prime_{\rm QCD} / \GeV)^2 C^{-1}$,
the relic density of stable glueballs will have no sigificant
gravitational effects, and if the metastable glueballs decay
well before recombination time ($\sim 10^{13} \second$), they
will not inject enough energy to observably disrupt BBN or the
CMB spectrum~\cite{Jedamzik:2006xz,Hu:1993gc,Slatyer:2012yq}.
However, an energy injection of $\gtrsim 10^{-10}$ of the DM
energy density can, depending on the injection time and channels,
have observational consequences if it occurs around recombination
time or later, either through CMB effects, or via cosmic ray
observations~\cite{Slatyer:2012yq,Dugger:2010ys}. Thus, it may be a
requirement that the meta-stable glueballs have lifetimes shorter than
$\sim 10^{13} \second$.

As an aside, note that the situation is different if $m_{b'}$ is light enough
that the lightest twin QCD states are mesons rather than glueballs.
The lightest meson state is a $0^{-+}$, so in the absence of lighter
twin sector states, will decay through higher dimensional operators,
as per the $0^{-+}$ glueball discussed below.
The difference is that its annihilations must now produce SM
final states, rather than purely twin-QCD states, so will
be much too suppressed to give it a sub-DM abundance. 
Number-changing interactions
(e.g. $3 \rightarrow 2$ processes) will reduce its density,
since other twin QCD states can decay to the SM, but since we are working
in the regime of
meson masses significantly larger than $1 \GeV$, the number-changing interactions
generally freeze out before the abundance can be reduced to sub-DM levels~\cite{Hochberg:2014dra}.
Thus, it appears that, in the absence of twin sector $CP$ violation and lighter
twin sector states,
the lightest $b'$ meson must decay before BBN time in order to avoid dangerous energy injection.
Since there are dimension-6 operators that could lead to this decay,
this condition is in principle easily satisfied.

We now turn to the question of the lifetimes of the metastable $0^{-+}$ and $1^{+-}$
glueballs mentioned in Section~II.   Let us first consider the case where there exist new SM--twin-sector
interactions in the UV completion of the TH theory that allow the $0^{-+}$ and $1^{+-}$
glueballs to decay.   For the $0^{-+}$ glueball the lowest dimension effective operators
conserving total $CP$ that allow the glueball to directly decay to the SM sector are of dimension 7, e.g.
operators of the schematic form $\bar{q}\gamma_5 q \times \tr(G^\prime \tilde{G}^\prime) /M^3$
(here $G^\prime$ are the $SU(3)^\prime$
field strengths, while $q$ stands in for SM fermions).  These lead to a lifetime,
$\tau_{0^{-+}} \sim 10^{-12} \ {\rm s}\ (M / 5\tev)^6 (3 \gev / \Lambda_{\rm QCD}^\prime)^7$,
irrelevantly short for astrophysical purposes (though interesting for displaced vertices at the LHC)
unless the scale suppressing the operator is raised
to $M \gsim 500 \tev$.   On the other hand, the leading effective operators for the $1^{+-}$
glueball are of dimension 10, e.g. operators of the schematic form
$\bar{q}\gamma_\mu\gamma_5 q \times \partial^\mu G^\prime G^\prime G^\prime/M_{}^6$.  These lead to a
lifetime $\tau_{1^{+-}} \sim 10 \ {\rm s}\ (M_{} / 5\tev)^{12} (3 \gev / \Lambda_{\rm QCD}^\prime)^{13}$, which is
of interest for indirect detection signals as discussed in Section~VII.

On the other hand, if only those interactions present
in the IR effective theory are considered then decays of the $1^{+-}$ glueball 
compatible with conservation of both angular momentum and $CP$ involve
combinations of on- and off-shell glueballs, which then decay to the SM via the Higgs
portal.  For example, $1^{+-}$ can decay to $0^{++}$
and an off-shell $0^{++}$ / $h$, in a $l=1$ state.  Since both $C$ and
$P$ are violated in this process, and the only interactions in the
twin sector that violate both $C$ and $P$ are $SU(2)^\prime$, the decay
of the $1^{+-}$ glueball needs to proceed through twin weak interactions
involving both axial and vector currents. We can estimate its decay
rate taking into account the leading heavy quark corrections to the
vector and axial currents, and assuming that the decay
to SM final states is via off-shell $0^{++}$ / $h$ mixing.
Using the (dominant) one-loop $b^\prime$ quark correction to the vector and axial
$SU(2)^\prime$ currents \cite{KaplanManohar,Juknevich}:
\begin{multline}
	\delta J_\mu^{\rm V} = \frac{g_3^{\prime 3}}{ 16 \pi^2 {m_{b^\prime}}^4} \partial_\alpha \ {\rm Tr}
		\left( \frac{1}{7} G_{\sigma \tau} \{ G^{\sigma \tau}, G_{\alpha \mu} \} \right. \\
		\left. - \frac{14}{45} G_{\mu \sigma} \{ G^{\sigma \tau}, G_{\tau \alpha} \} \right)
\end{multline}
\begin{equation}
	\delta J_\mu^{\rm A} = \frac{g_3^{\prime 2}}{ 48 \pi^2 {m_{b^\prime}}^2} \epsilon_{\mu \rho \tau \sigma}
		 {\rm Tr} (G^{\alpha \rho} \partial_\alpha G^{\tau \sigma} + 2 G^{\tau \sigma} \partial_\alpha G^{\alpha \rho}),
\end{equation}
the decay rate of the $1^{+-}$ glueball is estimated to be (ignoring dimensionless numbers and assuming $m_h^2 \gg m_0^2$),
\begin{equation}
	\Gamma_{1^{+-}} \propto \frac{v^2 \Gamma_{h}(m_0^*)}{m_{b^\prime}^{12} m_{Z^\prime}^{4} m_h^4 f^4} m_0^{22}
\end{equation}
and since $\Gamma_{h}(m_0^*) \sim m_0^* \sim m_0$ we see that $\Gamma_{1^{+-}} \propto m_0^{23}$.  Thus the decay rate of the $1^{+-}$ glueball depends very strongly on the mass of the glueballs, or equivalently
on the twin confinement scale.  For example, for $f/v=3$, 
\begin{equation}
	\tau_{1^{+-}} \sim \left( \frac{m_{b^\prime}}{15\gev} \right)^{12} 
	\times \left\{
	\begin{array}{l} 
		10^9 \ {\rm s} ~{\rm for}~\Lambda^\prime_{\rm QCD} \approx 1.5 \gev \\
		10^2 \ {\rm s} ~{\rm for}~\Lambda^\prime_{\rm QCD} \approx 3 \gev
	\end{array} \right.
\label{eq:DecayTime1+-}
\end{equation}
i.e. changing $\Lambda^\prime_{\rm QCD}$ by a factor of 2 leads to
a 7-order of magnitude difference in the lifetime of the $1^{+-}$
glueball! All that can be said is that, for reasonable values of the
parameters of the model, it appears that the $1^{+-}$ glueball may be a
long lived state. Since it ultimately decays to SM states, even a tiny freeze-out density
of $1^{+-}$ glueballs may be dangerous, specifically if the $1^{+-}$
lifetime is close to or after recombination time. 
 We therefore from now
on assume that, in the case where effects from higher dimensional operators can be neglected,
$\Lambda^\prime_{\rm QCD}$ and $m_{b^\prime}$ are such
that $\tau_{1^{+-}}\ll t_{\rm CMB} \simeq 10^{13} \ {\rm s}$, preferring larger values 
of $\Lambda^\prime_{\rm QCD}$ or smaller values of $m_{b^\prime}$. 

We finish by emphasising that the stability or duration of metastability of the $0^{-+}$ and $1^{+-}$ 
glueballs depends sensitively on whether the twin neutrinos are heavy or not, the mass scales $m_{b^\prime}$
and $\Lambda_{\rm QCD}^\prime$, whether twin $CP$ is conserved, and finally the nature of the interactions in the UV
completion between the twin and SM sectors in addition to the Higgs portal that might allow the glueballs to more
quickly decay.

\section{\label{sec:id}  Indirect Detection}

As well as leading to DM-SM scatterings in direct detection experiments,
the Higgs mixing portal between the twin and SM sectors may result in
SM energy injections from astrophysical DM-DM annihilations, leading
to potentially observable indirect detection signals. For the twin DM
candidates considered in this paper, annihilations proceed to lighter
twin sector states. For $\tau'$ and $\nu'$ DM, the dominant annihilation
channel is to a $b' \bar{b'}$ pair via $s$-channel $Z'$ exchange, due
to the colour factor enhancement of the final states available. Since
$SU(3)'$ is confining, the $b' \bar{b'}$ state will fragment into some
number of twin glueballs and mesons --- in the heavy-quark limit, we
would expect dominantly glueballs.

As previously discussed, if the $\nu'$ are
light, then most of the glueballs will decay down to the lightest
$0^{++}$ state by $\bar{\nu'} \nu'$ emission, with the $0^{++}$ state
then decaying via mixing with the SM Higgs. If there are no light hidden
sector states, then some of the glueballs may be metastable, while
others will decay via the SM Higgs. In either case, we expect some
combination of invisible decay products --- $\nu'$ or stable glueballs
--- and of off-shell Higgs particles, $h^*$. The latter will have some distribution of
mass squared values determined by the glueball masses and mass splittings. 

For $\Lambda_{\rm QCD}'$ greater than a few GeV, most of these $h^*$ should
be above the $b\bar{b}$ threshold. In that case, the SM injection products
will, to a good approximation, be a spectrum of $b\bar{b}$ with energies
determined by the fragmentation process.
For DM of
mass $m_{\rm DM} \gsim 63 \GeV$ as considered here,
the most sensitive probes of such astrophysical energy injection are
cosmic ray antiprotons and gamma rays. The most up-to-date constraints
on antiproton injection from dark matter annihilations come from the
AMS-02 experiment~\cite{AMS02,Giesen:2015ufa}. This sees an antiproton spectrum which is
higher, at energies $\gtrsim 20 \GeV$, than expected in \emph{some} models
of astrophysical secondary production and propagation. If one takes
these observations as disfavouring those models, and places constraints on
the DM antiproton contribution assuming the secondary production models
that better fit the AMS-02 data, one obtains constraints such as those
of~\cite{Giesen:2015ufa}, which finds that DM annihilating to $b\bar{b}$
with a thermal freeze-out cross section is disfavoured below $m_{\rm DM}
\simeq 100 \GeV$. If instead one includes the lower-secondary-production
models in one's uncertainty, the upper bound on the DM cross-section
increases by almost an order of magnitude, and DM annihilating to $b\bar{b}$
is still viable in the entire mass range (though at the lower end, necessarily
contributing significantly to the observed AMS-02 spectrum).

The best gamma-ray constraints on DM annihilation to hadrons, in this
mass range, come from the FERMI LAT instrument. In particular, a recent
FERMI analysis of dwarf spheroidal galaxies~\cite{Ackermann:2015zua} places
strong constraints on DM annihilating to $b\bar{b}$ with a thermal
freeze-out cross section, disfavouring NFW-profile DM in these galaxies
with $m_{\rm DM} < 100 \GeV$. Taking into account uncertainties
as to the DM profile in these galaxies, and/or using
the limits from the better-constrained diffuse Milky Way
halo~\cite{Ackermann:2012rg}, relaxes these constraints,
disfavouring $m_{\rm DM} \lesssim 20 \GeV$.

In our case, the uncertainty in the spectrum of $h^*$ energies, and so
$b\bar{b}$ energies produced, as well as the uncertainty regarding the
invisible decay fraction, mean that we cannot make precise predictions
for comparison with published limits. However, both the softening of the
$b\bar{b}$ injection spectrum, and the invisible decay fraction, will
act to decrease the strength of these constraints relative
to straightforward $b\bar{b}$ injection, and in combination with
the astrophysical uncertainties, this means that these models are 
not ruled out by current data. However, especially at low tuning, and so
low $f/v$, they predict indirect detection signals
potentially within the range of future observations.

Particularly interesting from the point of view of future measurements
are the tentative hints of a GeV-scale gamma-ray excess from the
Galactic Centre (GC)~\cite{Hooper:2013rwa, Abazajian:2014fta} in FERMI
data, which have now been corroborated by the FERMI collaboration
itself~\cite{Murgia}. This excess has a spectral and spatial profile
compatible with $\sim 40 \GeV$ DM annihilating to $b\bar{b}$, with
approximately thermal freeze-out cross section (or with $\sim 10 \GeV$
DM annihilating to leptons). While our DM masses are somewhat higher,
the softer spectrum expected may plausibly give a good spectral fit.
The higher mass and invisible decay fraction of our model would have
some difficulty replicating the approximately-thermal annihilation
cross section derived in current analyses, but systematic astrophysical
uncertainties are such that this may decrease with future observations.
Alternatively, if $\Lambda_{\rm QCD}'$ is low enough that a substantial
fraction of the $h^*$ produced have mass-squared below the $b\bar{b}$
threshold, then injection to $\tau^+ \tau^-$ and to lighter quarks may
be important, likely lowering the required annihilation cross
section (e.g.~\cite{Cline:2015qha}). The potential compatibility
with this existing signal provides additional motivation for better
understanding the twin QCD fragmentation process.

Another interesting possibility is that the fragmentation process
produces an appreciable number of metastable glueballs,
which may have a long enough lifetime to travel for astrophysically-relevant
distances before decaying. For this effect to be significant in galactic
DM observations, the lifetime of the glueballs would need to be
on kiloparsec ($\simeq 10^{11} \second$) scales (just
shorter than the lifetimes constrained by the CMB power spectrum).

A different and more striking signal may occur for smaller
lifetimes --- if the lifetime is $\gtrsim$
the radius of the Sun, $R_\sun \simeq 2 \second$,
but not much greater than the Earth-Sun distance, then metastable glueballs produced by annihilations
of captured DM in the Sun can escape before decaying, and their
decay products could be detected in charged cosmic rays or 
gamma rays~\cite{Batell:2009zp,Schuster:2009au,Schuster:2009fc,Ajello:2011dq}.
The solar capture rate of DM scattering spin-independently (and
momentum-independently) from SM nuclei is, taking a fiducial mass
of $m_{\rm DM} = 100 \GeV$, $C_\sun \simeq 10^{20} \second^{-1} \frac{\sigma_{\rm SI}}{10^{-45} \cm^2}$~\cite{Gondolo:2004sc} --- once enough
DM has accumulated in the Sun, an equilibrium will be reached in which
the annihilation rate is half of the capture rate, $\Gamma_{\rm ann} = \frac{1}{2}
C_\sun$. Since we have a precise prediction
for the scattering rate off SM nuclei, for given DM mass, 
this determines the annihilation rate in the Sun (for
annihilation cross sections of the magnitude
we are considering, the equilibrium rate is reached in much shorter than the lifetime
of the Sun). 
Ref. \cite{Schuster:2009au} considers the decay of metastable DM annihilation products
into leptonic final states, finding that FERMI gamma-ray observations
place an upper bound of $\Gamma_{\rm ann} f_{\rm dec} \lesssim 5 \times
10^{19} \second^{-1}$, where $f_{\rm dec} = e^{-R_\sun / \gamma \tau} - e^{-{\rm AU} / \gamma \tau}$ is the fraction of events, for a mediator of proper lifetime $\tau$
with boost factor $\gamma$, in which the decay occurs between the Sun and
the Earth. This is in the scenario
where DM of mass $100 \GeV$ annihilates inside the Sun to two mediator
particles, each of which then decay to two leptons. In our case,
only some fraction of the DM annihilation energy will go to metastable
glueballs, and the gamma-ray spectrum produced by each of these
will generally be softer, so we would expect the constraints to be less severe.
However, for lifetimes $\sim 1 \second$, but less than a few times $10^3
\second$, there is some possibility that future observations could place
limits on this scenario, or see a signal. Additionally, in the same way
that constraints on an excess of high-energy electron/positrons from the
Sun have been used to set limits on scenarios with leptonic decays
of the mediators~\cite{Ajello:2011dq,Casaus:2014cka,Mikhailov:2015vha}, a similar
analysis applied to antiprotons may provide stronger constraints on
the Higgs-portal models considered here.

\section{\label{sec:relativisticdof}  Equilibration of sectors} 


If the $\nu'$ are light, then we do not expect any of the glueballs to have
cosmologically relevant lifetimes. Also, for most
glueballs the decays involving $\nu'$ will occur at a faster rate
than decays involving the SM.
Since $\nu'$ interactions
with SM states are very suppressed, this raises the possibility that
most of the entropy in the twin QCD sector may be transferred to the $\nu'$,
and remain there decoupled from the SM. As discussed below, if the $\nu'$
are light, then they must be effectively massless as regards early-universe cosmology.
Still, they could potentially give a 
large DR contribution, which would be in conflict with
cosmological constraints on the extra number $\Delta N_{\rm eff}$
of effective neutrino species.

The minimum $\nu'$ energy density we might get would arise
from the $\nu'$ being in equilibrium with the SM until after the twin QCD
phase transition. In this case, since there were $g_* \simeq 75$ effective
relativistic degrees of freedom in the SM bath, and SM neutrinos
decouple from the SM bath when $g_{*, \rm SM} = 10.75$, we
have $T_\nu \simeq \left(\frac{75}{10.75}\right)^{1/3} T_{\nu'} = 1.9 T_{\nu'}$
at late times, assuming that the $\nu'$ are light enough to behave as DR.
Writing the energy density in DR at around CMB times as $\rho_{\rm
DR} \equiv (3 + \Delta N_{\rm eff})\rho_{\nu,{\rm SM}}$ (where
$\rho_{\nu,{\rm SM}} \equiv \frac{7}{8}\left(\frac{4}{11}\right)^{4/3}
\rho_\gamma$ is the naive SM neutrino energy density), the $\nu'$ give a
contribution to the effective number of neutrino species of
$\Delta N_{\rm eff} \simeq (1.9)^{-4} = 0.075$. If the $\tau'$ are also
light, the contribution is doubled.  
Non-instantaneous decoupling of the SM neutrinos gives
a contribution of $\Delta N_{\rm eff} = 0.046$ from the SM alone.
Planck has measured $\Delta
N_{\rm eff} = 0.15 \pm 0.2$~\cite{Ade:2015xua} (the $\pm 0.2$
being a simplified summary of a rather complex set of constraints:
see~\cite{Ade:2015xua} for details), so such a twin sector contribution
is entirely compatible with present-day constraints. However, future
observations, including improved astrophysical determinations of
the Hubble constant~\cite{Ade:2015xua}, large-scale structure
surveys~\cite{Hannestad:2014voa}, and a future cosmic-variance CMB
polarisation mission~\cite{Bashinsky:2003tk}, may be able to measure
$N_{\rm eff}$ to an accuracy of $\sigma(N_{\rm eff}) \sim 0.05$,
potentially testing the presence of a twin $\nu'$ bath.

Lowering the temperature at which the phase transition occurs
will lower the number of effective degrees of freedom in the SM bath, 
so will raise the contribution to $\Delta N_{\rm eff}$, up to $\simeq 0.1$
for temperatures just above the SM QCD phase transition. Below the SM
QCD phase transition, the \emph{minimal} contribution becomes $\simeq
0.5$, which is observationally marginal. However, we will see below
that we need to take $\Lambda_{\rm QCD}' \gtrsim 2.5 \GeV$ for
sufficiently fast equilibration in any case, which is well above
$\Lambda_{\rm QCD}$.

If the $\nu '$ are too heavy to behave as DR until CMB times, 
they will constitute an extra (possibly hot) DM abundance.
For $m_{\nu'}$ greater than a few keV, and less than weak scale values, the $\nu'$ will constitute a
warm/cold DM abundance with far too high a relic density.
If $m_{\nu'}$ is smaller than that, but still greater than $\mathcal{O}(10 \eV)$,
then they will constitute too large a warm/hot DM abundance (e.g.~\cite{Viel:2007mv}).
So, if $m_{\nu'} < m_h/2$, it is required to be less than a few eV, the same
applying to $\tau'$.

Thus, if the $\nu'$ are light, then they are certainly much lighter
than the glueball mass splittings. In this case, as mentioned above,
all of the glueballs are unstable, either against decays involving
$\bar{\nu'} \nu'$ or $h^*$. The question of interest then becomes
how much energy density from the twin QCD bath ends up in the
eventually decoupled $\nu'$ and SM baths, and whether the $\nu'$
represent an observationally significant DR abundance.

In the event that all of the entropy density of the twin QCD bath
ends up in the $\nu'$, we can estimate the resulting contribution
to $\Delta N_{\rm eff}$ as above.
If $m_{b'} \gtrsim 5 \GeV$, then during
the first stages of its Boltzmann suppression, the $b'$ coupling to the SM
is large enough to equilibrate the SM and twin gluon sectors.
Thus, the scenario resulting in maximum $\nu'$ energy density is that in which
the SM and twin QCD sectors decouple shortly after the $b'$
go non-relativistic. In this case, the entropy density of the gluon
bath, corresponding to $\sim 16$ effective relativistic dof,
stays within
the twin sector. Since all of this gets transferred eventually to the $\nu'$,
then if (as expected) they have a quasi-thermal spectrum,
they will have a temperature 
$\sim \left((16 + 2 \times
\frac{3}{4})/(2 \times \frac{3}{4})\right)^{1/3} \simeq 2.3$ times the
$a^{-3}$ scaled SM
temperature before the phase transition.
So, they contribute $\Delta
N_{\rm eff} \simeq (2.3 / 1.9)^4 \simeq 2.0$. 

A $\Delta N_{\rm eff}$ contribution this large is \emph{incompatible}
with the existing cosmological observations.
In fact, even if only the entropy of the twin QCD bath just about
the critical temperature is transferred to the $\nu$', then
the resulting $\Delta N_{\rm eff} \simeq 0.4$ is in tension
with observations.
It is therefore a requirement that most of the entropy in the unconfined
phase of the
QCD bath
is transferred to the SM, rather than to the
$\nu'$. Logically, this could occur in a number of ways:
\begin{itemize}
\item If twin QCD interactions with the $\nu'$ and SM are too slow
to transfer significant entropy to either in the first Hubble time, then
the only process that can become important is glueball/meson decays.
Then, the energy transferred to the $\nu'$ and SM is set by the overall
branching ratio of these decays.

However, in our scenarios, $SU(2)'$ mediated processes are generally faster
than Higgs portal ones, and we expect the branching ratio to favour
$\nu'$ over SM, so this does not solve the problem. 

\item If twin QCD interactions with both the $\nu'$ and the SM are
fast enough, then $\nu' \leftrightarrow {\rm QCD}' \leftrightarrow {\rm SM}$
interactions may be fast enough to bring the $\nu'$ and SM sectors
(almost) to equilibrium, and so
to (almost) restore the $\Delta N_{\rm eff} = 0.075$ situation derived
at the beginning of this section.  (Note that direct interactions between $\nu'$
and SM states are extremely suppressed, since diagrams must involve
$SU(2)'$ gauge bosons and multiple Higgs's. The $\nu'$ may have a small
interaction with the twin Higgs, giving them a $\lesssim \eV$ mass, but
this coupling is again far too small to bring about equilibrium.)
\end{itemize}

It is not trivial for interactions during the confined phase to be
fast enough, since the glueball bath is non-relativistic, and its
equilibrium number density falls very fast as the temperature drops. 
Energy transfer from the glueball bath to the SM needs to be fast enough
that it is still operating until the SM and $\nu'$ have reached almost
the same temperature. 

\begin{figure}[h!]
  \includegraphics[scale=0.53]{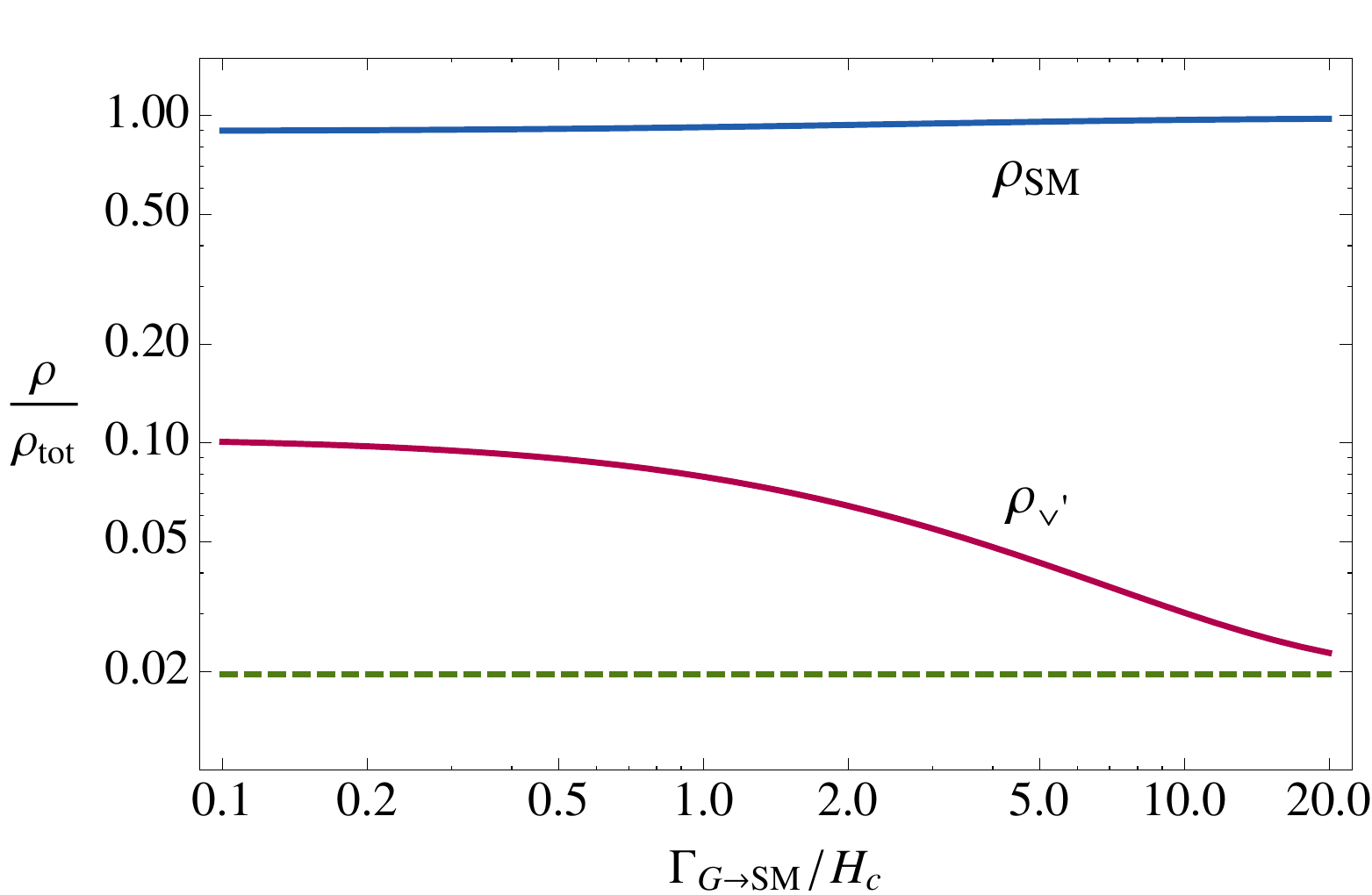}
 \caption{\label{fig:rhoG} Relative energy density contributions,
after $SU(3)'$ phase transition, of SM (blue) and $\nu'$ (red) baths,
for given glueball-SM energy transfer rate $\Gamma_{G \rightarrow {\rm SM}}$
(compared to Hubble rate at start of phase transition $H_c$).
This calculation assumes that the SM, ${\rm QCD}'$ and $\nu'$
baths are in thermal (chemical) equilibrium at $T=T_c$, which
will not be the case if the coupling is weak enough (that is,
the small-$\Gamma$ part of this graph would have $\rho_{\nu'}$
even larger in a realistic scenario).
The dashed green line shows the number equilibrium density of $\nu'$.
}
\end{figure}

In the confined phase, the fastest processes transferring energy
${\rm QCD}' \rightarrow$ SM are glueball and meson decays,
for example, the $0^{++} \rightarrow h^* \rightarrow$ SM decay
discussed previously, and possibly SM-${\rm QCD}'$ scatterings. Parameterising the effective SM decay rate of the
glueball bath as $\dot\rho_G = - \Gamma_{G \rightarrow {\rm SM}} \rho_G + \dots$,
and solving the Boltzmann equations numerically shows that 
$\Gamma_{G\rightarrow {\rm SM}} \gtrsim 10 H_c$ is required to significantly
reduce the $\nu'$ abundance, where $H_c$
is the Hubble rate at the start of the phase transition -- see Figure~\ref{fig:rhoG}.
Since $H \propto T^2$, and the decay rates go as higher powers 
of $\Lambda_{\rm QCD}'$ (e.g. $\Gamma_{0^{++}} \propto \Lambda_{\rm QCD}'^7$), increasing
$\Lambda_{\rm QCD}'$, and so $T_c$, makes it easier to fulfil this condition,
which is satisfied for $\Lambda_{\rm QCD}' \gtrsim 2.5 \GeV$ for our fiducial parameters.
Since the decay and scattering rates of most of the glueball and meson
states have not been calculated, the precise bound is uncertain. However,
because these rates increase as much higher powers of the mass scale
than the Hubble rate does, such uncertainty in the rates at a given energy
has a small effect on the $\Lambda_{\rm QCD}'$ bound.

The ${\rm QCD}' \rightarrow$ SM energy transfer from a mixture of
confined and unconfined phases, at the critical temperature, should (by
detailed balance) be at most that from a wholly-confined-phase mixture
(assuming that the confined phase loses most energy from processes with
only twin sector particles in the initial state). Figure~\ref{fig:rhoG}
assumes that the rate is that of the confined phase decay approximation,
from the critical temperature downwards. So, either $\Lambda'_{\rm QCD}
\gtrsim 2.5 \GeV$, in which case we are driven almost to equilibrium,
giving a $\Delta N_{\rm eff}$ contribution of $\simeq 0.075$, or else
a large fraction of the entropy of the twin QCD bath goes into the
$\nu'$, giving a $\Delta N_{\rm eff}$ contribution of $\sim 0.4 - 2.0$
in conflict with observations.

\section{\label{sec:summary} Summary and Conclusions}

We have considered the status of DM (\emph{without} a matter-antimatter
asymmetry) in models based on the TH mechanism
\cite{TwinHiggs}\cite{Chacko2005,ChackoNomura,Barbieri2005}, focusing
on its simplest implementation, the Fraternal Twin Higgs
\cite{FraternalTH}, that contains just $SU(3)'$ and $SU(2)'$ gauge
interactions and only a third generation of fermions. 
While such models may feature a first-order $SU(3)'$ phase
transition at temperatures around or below DM freeze-out,
this transition does not result in significant entropy production.

The most natural DM candidate is the twin tau lepton $\tau'$, which
requires masses $m_{\tau'} > m_h / 2$ in order to provide the observed
DM abundance. In particular, we find $m_{\tau'} \approx 63 - 130 \ \gev$
for ratios $f/v = 3 - 5$, which implies a very mild $20-8 \%$ tuning. DM
scattering off SM nuclei happens via Higgs exchange, and direct detection
signatures lie below LUX current bounds \cite {LUXresults} but within
LUX future sensitivity \cite{LUXprojection}. A very natural possibility
within the Fraternal Twin Higgs scenario is that of multicomponent
DM. In particular, DM made of both $\tau'$ and $\nu'$ species arises
naturally as soon as the masses of both states are of similar size,
though the phenomenology of this scenario is very similar to that of
single-$\tau'$ DM, modulo the lack of any $\Delta N_{\rm eff}$
contribution. In the case where $m_{\tau '} + m_{\nu '} > m_{W'}$,
$W'^\pm$ gauge bosons become stable and may significantly contribute to
the DM density, leading to a scenario where DM is made of three
different species of twin particles ($\tau'$, $\nu'$ and $W'^\pm$). We
find, however, that this three-component scenario requires large values
of $f/v$, that result in fine-tuning between $5 - 1 \%$, substantially
worse than the single- or two-component cases. We briefly mention the
possibility of twin baryon DM, where the DM particle would be a $\Delta
'$ baryon made of three $b'$ quarks. The efficient annihilation of
$\bar {b'} b'$ pairs via twin strong interactions requires extremely
large quark masses if the observed DM abundance is to be reproduced. In
particular, we find that masses $m_{b'} \gtrsim 1 \tev$ are required,
which translates into $f/v \gtrsim 30$ for $y_{b'} \approx 0.2$, and
therefore a fine-tuning worse than $0.5 \%$.

Regarding possible indirect detection signatures, annihilation of DM
particles happens mainly to $\bar {b'} b'$ pairs, which mostly results
in glueball final states as a result of the fragmentation process in
the twin QCD sector. Glueballs will decay both into SM final states and
into invisible twin sector states (either to $\bar {\nu '} \nu'$ in the
case of light $\nu'$, or to (meta)stable twin glueballs). The former
will consist mostly of $\bar b b$ pairs, giving broadly WIMP-like indirect
detection phenomenology, though with a presently non-calculable
injection spectrum. The masses of our DM candidates are large enough
that there is no inconsistency with current data, though
future observations should probe relevant regions of parameter space.
Decays of metastable glueballs may also have cosmological / astrophysical
consequences, though the strong dependence of these lifetimes
on $\Lambda_{\rm QCD}'$, and on the UV physics, means that
precise predictions are not possible.

In the case where $m_{\nu '}$ is small, a non-zero contribution to the
number of effective neutrinos, $\Delta N_{\rm eff}$, is a prediction
of the theory. This contribution depends on the value of the twin
confinement scale and we find that for $\Lambda'_{\rm QCD} \gtrsim 2.5 \
\gev$, the $\nu'$ bath remains in equilibrium with the SM bath after
the twin QCD phase transition and $\Delta N_{\rm eff} - \Delta N_{\rm eff, SM} \approx 0.075$,
a value compatible with current bounds but within reach of future
measurements. If the $\nu'$ and SM baths fell out of equilibrium before
the twin QCD transition, all the energy in twin glueballs is damped
into the $\nu'$ bath, in which case we would find $\Delta N_{\rm eff}
- \Delta N_{\rm eff, SM}
\approx 2$, strongly disfavoured by current bounds. 
In the case of heavy twin
neutrinos ($m_{\nu'} > m_h / 2$), there are no light twin sector
states, so no contribution
to $\Delta N_{\rm eff}$.

We remark that, independent of twin sector DM opportunities, our
analysis highlights cosmological and astrophysical constraints on TH
related models. The $\Delta N_{\rm eff}$ bounds are independent of whether there
is a twin sector DM abundance, while the relic density calculations of
Sections~\ref{sec:tauDM} and~\ref{sec:multicomponentDM} demonstrate
that, for some mass parameters, the minimal TH model naively produces a
super-DM abundance of stable states, which would need to be reduced by
introducing new decay operators or hidden sector states.

Finally, we note that similar investigations of DM in Twin Higgs models
have been carried out by other groups \cite{Craig:2015xla} and \cite{Farina:2015uea}.

\begin{acknowledgments}
We wish to thank Nathaniel Craig, Roni Harnik, Matthew McCullough \& the participants
of the CERN Neutral Naturalness workshop where this work was first presented for useful discussions. We also thank Subir Sarkar and Mike Teper for helpful conversations.
IGG \& JMR are especially grateful to Kiel Howe for many discussions of Twin Higgs theories.
RL thanks the Berkeley Center for Theoretical Physics for hospitality during completion of this work, and JMR
thanks the CERN Theory Group for hospitality.  IGG is financially supported by the STFC and a Scatcherd European
Scholarship from the University of Oxford.  RL acknowledges financial support from an STFC studentship.
\end{acknowledgments}


\bibliography{twinDM} 

\end{document}